# Ultrathin and Multicolor Optical Cavities with Embedded Metasurfaces

*Amr M. Shaltout [1, 2, †], Jongbum Kim [1, 3, †], Alexandra Boltasseva [1],*

*Vladimir M. Shalaev [1], and Alexander V. Kildishev [1, *]*

School of Electrical & Computer Engineering and Birck Nanotechnology Center, Purdue University, West Lafayette, IN 47907, USA

Geballe Laboratory for Advanced Materials, Stanford University, Stanford, California 94305, USA

Institute for Research in Electronics and Applied Physics, University of Maryland, College Park, Maryland 20742, USA

**Abstract:**

Over the past years, photonic metasurfaces have demonstrated their remarkable and diverse capabilities for achieving advanced control over light propagation by confining electromagnetic radiation within the deeply subwavelength thickness of these artificial films. Here, we demonstrate that metasurfaces also offer new unparalleled capabilities for decreasing the overall dimensions of integrated optical components and systems. We propose an original approach of embedding a metasurface inside one of the most fundamental optical elements – an optical cavity – to drastically scale-down the thickness of the optical cavity. We apply this methodology to design and implement a metasurfaces-based nano-cavity where the Fabry-Pérot interferometric principle has been modified to reduce the cavity thickness below the conventional λ/2 minimum. In addition, the nano-cavities with embedded metasurfaces can support independently tunable resonances at multiple bands. As a proof-of-concept, using nano-structured metasurfaces within 100-nm nano-cavities, we experimentally demonstrate high spatial resolution color filtering and spectral imaging. The proposed approach can be extrapolated to compact integrated optical systems on-a-chip such as VCSEL's, high-resolution spatial light modulators, imaging spectroscopy systems, and bio-sensors.

**Introduction:**

Photonic metasurfaces [1-6] have enabled a new paradigm of controlling light by passing it through planar arrays of nano-structured antennas to induce a swift change of phase and/or polarization. Consequently, photons propagating through metasurfaces could be 'processed' to undergo a change to their momentum, angular momentum, and/or spin state. This has led to a relaxation of Snell's law [2], a pivotal relation in optics, and has enabled an entirely new family of flat optical elements. These elements can provide diverse functionalities for a plethora of applications including light bending devices [2, 7], flat lenses [8-11], holograms [12-17], wave-plates, [18-22] as well as devices with chiral [23-28] and bianisotropic [29, 30] optical response. A typical metasurface generates its functional output in the far field, which requires a separation distance between the metasurface and the location of the detected output. Therefore, even though a given metasurface could be ultrathin, an integrated optical system that requires cascading of the metasurface with another optical component may not fully utilize its capability to reduce the overall dimensions of the system.

In this work, we embed metasurfaces inside a most fundamental optical element - an optical cavity - to substantially reduce its thickness (up to 2 times). Such significant reduction in the transverse dimension is achieved without changing the optical mode for the sake of cascading and integration with other optical devices. Optical cavities are the major components of lasers and other numerous interferometric optical devices; hence, minimizing the cavity dimensions below the diffraction limit is the key for many applications, including realization of nano-lasers [31-33], and the enhancement of spontaneous emission rate due to the Purcell effect [34]. This enhancement, which is useful for single photon sources [35, 36] and low threshold lasing [37-39] and is inversely proportional to the volume of the cavity, has motivated researchers in nanophotonics in quest of solutions minimizing the cavity dimensions [40-44].

In the following sections, we present a metasurfaces-based Fabry-Pérot nano-cavity arranged of two parallel reflective metal layers and a metasurface sandwiched between them. First, we theoretically study the impact of the metasurface on the resonant condition of nano-cavity based upon the interferometric principle. Then, we discuss the structure fabrication and the experimental results showing that a 100-nm nano-cavity supports resonances within the wavelength band from 500 to 800 nm and demonstrate successful application of the fabricated nano-cavity samples as on-a-chip ultrathin color filters. In addition to the reduced cavity thickness, the metasurface can also add another degree of freedom to controlling the cavity resonance. Now it can be achieved through adjusting the parameters of the metasurface – not only by adjusting the cavity thickness. This is especially useful for building multiple cavities of the same thickness on a planar chip resonating at different wavelengths. These additional degrees of freedom also allow us to obtain dual-band resonances or even produce colored images by using a single planar nano-cavity of the same thickness by just choosing a proper design for the internal metasurface.

**Results:**

*Theoretical Formulation:*

The resonant condition of a conventional Fabry-Pérot cavity made of two reflecting mirrors occurs when the round trip optical phase inside the cavity is an integer multiple of $2\pi$. This imposes a limitation on the cavity thickness to accumulate the required phase. Our approach to overcome this limitation is to place a metasurface inside the two mirrors which induces a strong phase shift and compensates for the reduced accumulated phase in the rest of the cavity.

Figure 1(a) is a schematic of the nano-cavity arranged of two silver layers as reflecting surfaces, an silver-based metasurface separated from the bottom Ag mirror by an alumina spacer layer, and a polymer for filling the rest of the cavity. Figures 1(b) and 1(c) demonstrate the cross section of a Fabry–Pérot cavity without and with the embedded metasurfaces, respectively. At the resonance, the round-trip phase-shift inside the cavity satisfies the following formulas:

$$\frac{4\pi L}{\lambda} = 2\pi m, \quad \text{without metasurface,}$$

$$\frac{4\pi L}{\lambda} + \varphi_{ms} = 2\pi m, \quad \text{with metasurface,} \tag{1}$$

where $L$ is the thickness of the cavity and $\varphi_{ms}$ is the phase-shift induced by the metasurface. The conventional formula implies that for the metasurface-free cavity, a minimum cavity thickness $L_{min}$ is on the order of $\lambda/2$ when $m = 1$. In the case of an embedded metasurface, the induced phase $\varphi_{ms}$ reduces the required phase $4\pi L/\lambda$ for the rest of the cavity, and hence enables $L_{min}$ to go below $\lambda/2$. The underlying physical mechanism that induces the strong phase shift by the metasurface is the excitation of gap-plasmon resonators [45-47] and slow plasmonic modes that propagate inside the alumina spacer layer between the bottom silver layer and the metasurface structure. The gap-plasmon resonators can create large phase shifts for the wavefronts propagating inside the cavitity.

The phase-shifts induced by the gap-plasmons can be controlled by adjusting the lateral dimensions of the metasurface Ag nano-strip layer. By increasing the nano-strip width, the gap-plasmons propagate over a greater lateral distance and acquire additional phase, thereby decreasing the minimum cavity phase-shift required for resonance [43].

*Implementation & Experimental Results:*

Figure 2(a) shows a cross-sectional schematic of the nano-cavity. It is implemented with 15-nm-thick Ag mirrors, 40-nm-thick Alumina spacer, 22-nm-thick Ag metasurface gratings, and a polymer thickness in the rest of the cavity of 60 nm, which makes the whole cavity thickness between the two mirrors to be around 100 nm. The structure has a lateral period of 150 nm. Figure 2(b) shows a $30^0$ tilted view of the nano-cavity where a cross-sectional shear is introduced through the upper Ag layer and the polymer layer till the metasurface layer. Five different samples are fabricated with different metasurface widths of $W$ = 21, 35, 45, 56, and 71 nm. All the samples are fabricated on the same chip, and hence, they hold the same cavity thickness and any difference they exhibit in their optical performance is attributed to the different width ($W$) in their metasurface structure. (See METHODS section for the detailed fabrication and measurement procedure).

To demonstrate the impact of the metasurface on the nano-cavity, spectral transmission measurements are conducted on the nano-cavities with and without the metasurface. Figure 2(c,d) demonstrates spectral measurements and the transmission data respectively, obtained from finite-element simulation with the incident polarization being perpendicular to the Ag metasurface gratings, since this polarization enables gap-plasmon excitation across the width of the meta-gratings [47]. Comparing the spectral data without a metasurface (dashed line) to those of metasurface-based nano-cavities, we conclude that the metasurface makes the resonant wavelength longer across the visible and NIR spectra. As previously discussed, increasing the metasurface Ag grating dimensions increase the induced phase shift inside the metasurface $\varphi_{ms}$, and according to Equation (1), increasing $\varphi_{ms}$ causes an increase in wavelength λ for the equation to be satisfied. Therefore, resonant wavelength increases as the Ag gratings width increases across the values $W$ = 21, 35, 45, 56, and 71 nm. All these samples resonating at different wavelengths are having the same cavity thickness. The structured metasurface adds a new degree of freedom of controlling resonant wavelength other than the cavity thickness. On the other hand, the impact on the metasurface on the resonant wavelength when light polarization is parallel to the gratings is not significant since this polarization is poorly coupled to the gap-plasmonic mode. This reflects in the spectral measurement and simulation data shown in Figure 2(e,f). The simulation data is well-matched with the experimental data; however, the simulated transmission data within the wavelength range from 400 to 500 nm are higher, since the utrathin Ge adhesion layer is excluded from the simulation domain [48].

*Dual-band Resonance:*

The formula of Equation (1) as well as the measurements in Figure 2(c) indicates that metasurface based nano-cavities are dependent on both the cavity thickness $L$ and the metasurface induce phase $\varphi_{ms}$, making the resonance controllable via two independent degrees of freedom. If the metasurface structure is designed such that it can induce a strong phase-shift $\varphi_{ms}$ at more than one wavelength, then multiple resonances can occur at different wavelengths that can be independently controlled through careful design of the metasurface. This enables a new technique of obtaining multi-band resonance other than having different higher modes standing waves across the cavity thickness (modes with $m > 1$).

To obtain a dual-band resonance, we implement a nano-cavity where the metasurface is composed of alternating grating widths of 35 and 56 nm as shown in Figure (3). We have had two samples, each of them has its grating width designed with only one of these two values. They were part of the samples whose results are shown in Figure 2. We demonstrate the individual response of each of these two samples in Figure 3 along with the sample implemented with the alternating grating. The sample with the hybrid Ag grating dimensions have demonstrated dual-band resonances which coincide with the two individual resonances obtained by the other two samples. This can be attributed to the metasurface's hybrid dimensions which will support gap-plasmonic modes at both wavelengths supported by grating widths of 35 and 56 nm. The small peak at a wavelength similar to the resonance of the cross-polarized light is attributed to the strong absorption of Ge layer within the wavelength from 400 nm to 500 nm (See the figure S2 in the supplementary document). The different resonant wavelengths obtained through this technique can be independently tuned by a separate adjustment of different grating widths.

*Ultra-thin Color Filtering and Imaging:*

One of the straightforward and promising applications of optical cavities is spectral filtering. The polarization sensitive color-filtering can be obtained from the metasurface embedded nano-cavity due to its narrow linewidth and broad tunability of resonance in the visible range. In addition, from white light through corresponding shaping of nano-cavities, color images can be generated in planar surface which make this device easy to be integrated with other optical components. With the same nano-cavity thickness of 100 nm as in previously shown results, we implement six different nano-cavities, each of them is shaped to one of the letters of the word PURDUE, with each letter having a dimension of 500 μm × 420 μm. The letters P, U, R, D, U are shaped with Ag metasurfaces having grating widths of $W$ = 21, 35, 45, 56, and 71 nm, respectively, which are the same dimensions of metasurfaces as demonstrated in Figure 2. Letter E is implemented using a metasurface with alternating gratings of 35 and 56 nm similar to the one in Figure 3. Broadband light from a white lamp is filtered through a polarizer, and applied to the sample, and multi-colored images are captured from the other side of the sample. Figure 4(a) demonstrates the color images obtained for different polarization directions of incident light with $0^0$ corresponding to perpendicular polarization, and $90^0$ to parallel polarization with respect to Ag gratings. The colored image is in good agreement with the spectral data in Figures 2-3. First, the part of the cavity outside the metasurface filters a sky-blue color that corresponds to its spectral data. The images of samples 1, 2, 3, and 4 for $0^0$ polarization have their resonant wavelengths pushed to the spectra of green, yellow, orange, and red, respectively. For sample 5, the spectral data is in the NIR outside the visible spectrum explaining the dark appearance of the letter E. For sample 6, there is a dual-band (such as the one in fig 3) which is a hybridization of two colors leading to the generation of the brown color for letter F. For all the six samples, the $90^0$-polarized light generates a blue color similar to the one in the metasurface-free background with a smaller intensity. This is also in agreement with the experimental data shown in Figure 2. For other polarizations between $0^0$ and $90^0$, the obtained spectrum is a linear combination of the horizontally and vertically polarized spectra. Figures 4(c,d) present the measured spectra for these polarizations for sample 4 and 6. In order to understand what color will be the result of this linear superposition, we need to use the standard 1913 CIE color map that maps the spectral data onto a corresponding color. This map exhibits the property that a linear superposition of two spectra will correspond to the linear superposition of their corresponding locations on the map (i.e, along the line connecting the two locations). The map location of the six samples for horizontal and vertical polarizations were obtained (The methods section explains the tools used to obtain them). The line connecting the two locations in the map should reflect the colors generated when the polarization is rotated between $0^0$ and $90^0$. Figure 4(a,b) indicates a good agreement between the color images and the corresponding position on the 1931 CIE color map.

Without using metasurfaces, this kind of multi-color filtering would require multi-cavities with varying thicknesses. The metasurface enables using the same cavity, while tailoring the color filter through adjusting the lateral dimension of the metasurface. Tuning the incident polarization adds another degree of freedom to light control. In addition, using alternating grating widths (like the dual-band) can also be beneficial to mix-up colors.

**Discussion:**

A unique methodology of utilizing metasurfaces to achieve miniaturized nanophotonic devices is introduced. Metasurface-based nano-cavities are proposed with dimensions that can go below the conventional diffraction limit of $\lambda/2$ in the material between the cavity mirrors. Embedding metasurfaces inside cavities provides other degrees of freedom for controlling resonant wavelengths in addition to adjusting the cavity thickness or dielectric material index. Several cavities are implemented on the same planar chip with the same thickness and we are able to obtain different resonant wavelengths through controlling the width of nano-structured metasurface elements. Multi-band filtering, coloration, and imaging is also made possible inside the same cavity by controlling the degrees of freedom provided by the metasurface. The proposed approach could have key impact for many optical applications including, but not limited to, nano-lasers and ultra-compact filters, as well as other interesting physical applications related to spontaneous emission enhancement based on the Purcell effect, cavity quantum electrodynamics, structural coloration, as well as plasmonics enhanced optoelectronic devices, including spatial light modulators with subwavelength resolution, spectroscopic imaging and bio-sensing systems.

**Methods**

**Sample Preparation:** The designed structure is compatible with nanofabrication techniques. 15-nm-thick Ag and 40-nm-thick $Al_2O_3$ films were subsequently deposited on the glass substrate by electron beam evaporator. To fabricate Ag metasurface (grating structure), positive electron beam resist (ZEP 520 A) was spin-coated and grating arrays were defined by electron beam lithography (Vistec VB6) followed by lift-off process. On top of the patterned resist, 22-nm Ag film was deposited with a 3-nm Ge layer used to improve adhesion, and the sample was dipped in ZDMAC (dimethylacetamide) for 20 minutes. For planar surface of spacer layer on top of metasurface, we spin-coated 800nm PMMA on top of metasurface and etched it down to desired thickness (600nm or 200nm) by reactive ion etching process (Panasonic E620 etcher). The roughness of etched PMMA spacer on top of metasurface is around 3 nm. Finally, 3-nm Ge film and 30-nm Ag film for top mirror layer and 40-nm alumina layer for protection of Ag were deposited on top of PMMA spacer.

**Sample Characterization:** The transmission spectra of fabricated nano-cavities were measured using the spectroscopic ellipsometer (J. A. Woollam Co., V-VASE). The light source is a xenon lamp with a broadband Visible to NIR spectrum. The diameter of the incident beam was set to be 400 μm. The beam sequentially passed through a monochromator, a polarizer, and it is used to expose the sample, and then it's collected by a detector. The color image was captured with simple optical set-up. The white lamp was used as light source and linear polarizer was placed between lamp and source. The image was captured with commercial camera in mobile phone (Samsung Galaxy S3) with 4X magnification. ISO was set to 100, auto-contrast was off, and the micro lens was attached to camera.

**Simulation:** we validated the experimental spectra with numerical simulations using a commercially available software based on the Finite Element Method (COMSOL Multiphysics, Wave Optics Module). 1913 CIE color map was calculated by Matlab with an open-access code from the Matlab website, https://www.mathworks.com/matlabcentral/fileexchange/29620-cie-coordinate-calculator.

**Acknowledgements:**

This work was supported by AFOSR grant 123885-5079396, ARO grant W911 NF-13-1-0226, NSF MRSEC grant 3002095871, and DARPA Extreme Optics and Imaging (EXTREME) program.

**Contributions:**

A.M.S. and J.K. developed the concept. A.M.S. designed and simulated the prototype structures. J.K. carried out the multi-layered fabrication of the devices. J.K and A.M.S. performed the experiments. A.V.K., A.B. and A.V.S. oversaw the project and provided the facilities. All the authors contributed and approved the manuscript.

**Competing financial interests:**

The authors declare no competing financial interests.

**Author Information:**

[†]These authors contributed equally to the manuscripts.

*E-mail: kildishev@purdue.edu

**References:**

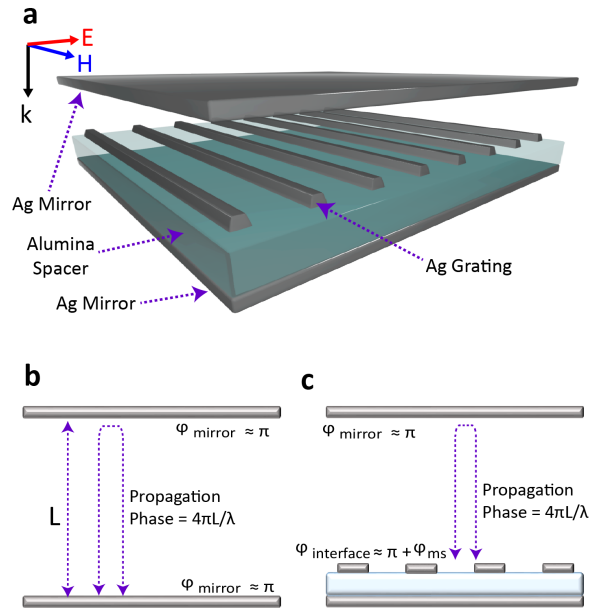

Fig 1: **(a)** Schematic of the metasurface based nano-cavity. **(b,c)** Comparison in phase and resonance conditions between: **(b)** conventional Fabry-Perot resonator that utilizes two parallel mirrors and **(c)** resonator with a reflecting metasurface placed in between the two mirrors.

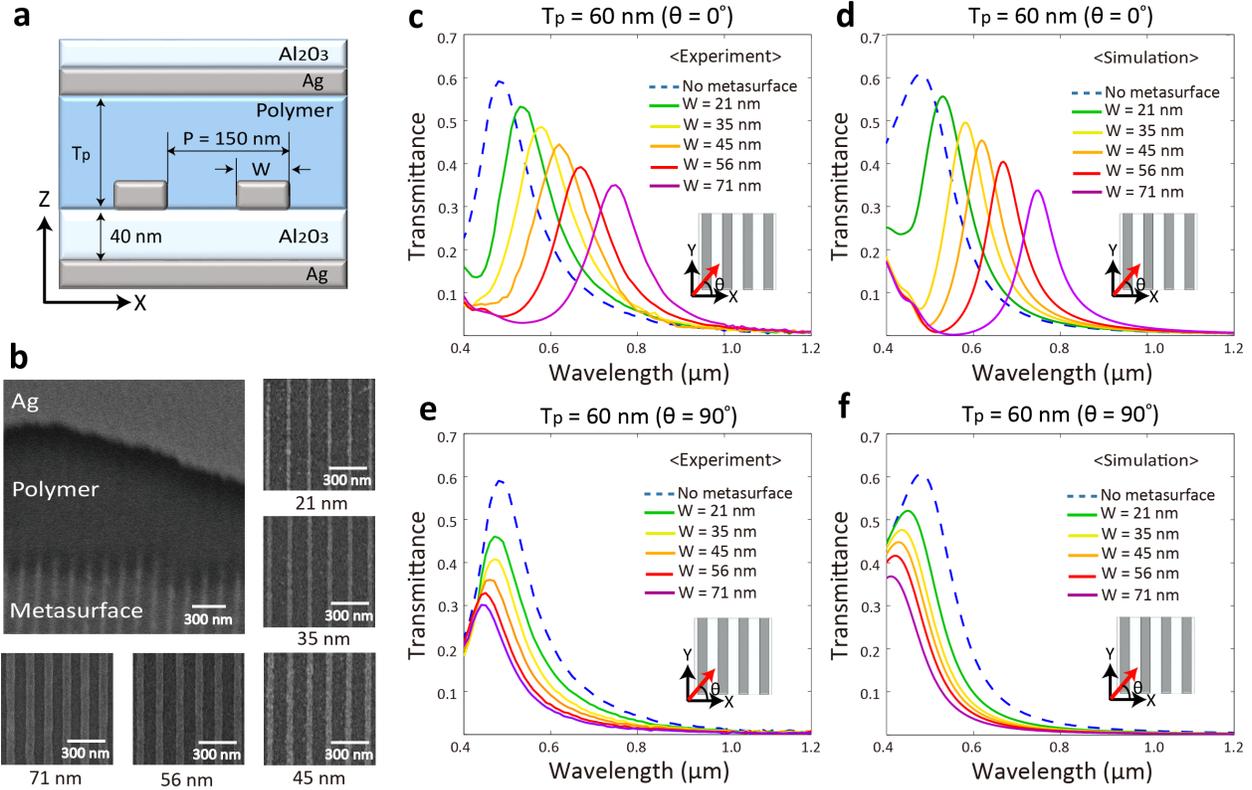

Fig 2: **(a)** Cross-sectional schematic of the metasurface based nano-cavity. P and W is the perodicity and linewidth of Ag grating, respectively. $T_p$ is the thickness of polymer layer. The thickness of both top and bottom Ag mirror is 15 nm and the thickness of grating is 22 nm. **(b)** 30° tilted view and Top view of FE SEM image of fabricated nano-cavity. Experiment (c,e) and simulation (d,f) of transmission spectra of nano-cavity with 60 nm thick polymer spacer (inset in (a): the schematic of polarization direction, θ is the angle of polarization of incident wave): (c,d) θ = 0° and (e,f) θ = 90°.

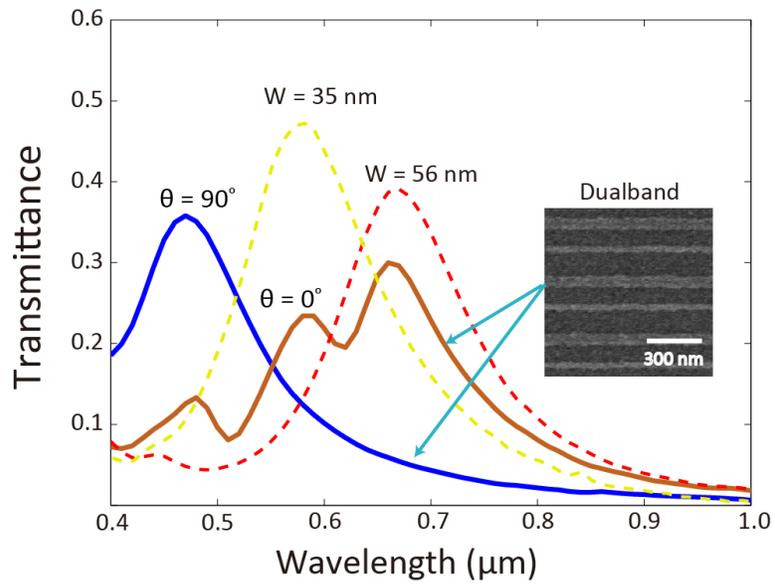

Fig 3: Experimentally obtained transmission spectra of dualband nano-cavity with 60-nm-thick polymer spacer: The dualband metasurface consists of two different Ag gratings with 35 nm and 56 nm linewidth.

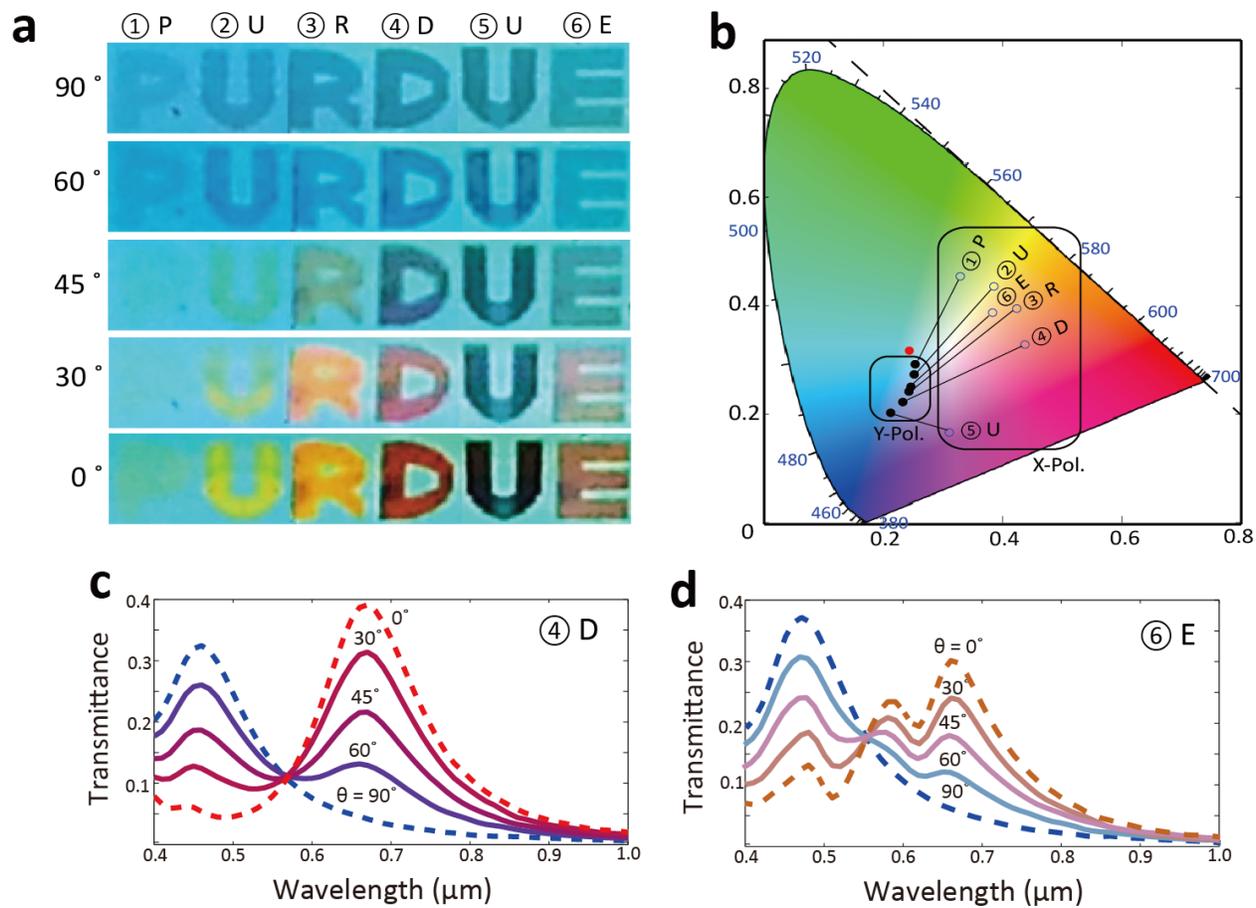

Fig 4: **(a)** Color image of the nano-cavity with different linewitdh. White light illuminates the sample via a linear polarizer. The angle of polarization is varied from 0° to 90°. (b) The variation of color with angle of polarization is analyzed with the CIE 1931 color mapping. The red dot represents the background color. (c,d) The transmission spectra of nano-cavity at different angle of polarization: (c) for sample 4 (letter 'D') and (d) for sample 6 (letter 'E').

# Supplementary Information Materials

# Ultrathin and Multicolor Optical Cavities with Embedded Metasurfaces

*Amr M. Shaltout [1, 2, †], Jongbum Kim [1, 3, †], Alexandra Boltasseva [1],*

*Vladimir M. Shalaev [1], and Alexander V. Kildishev [1, \*]*

School of Electrical & Computer Engineering and Birck Nanotechnology Center, Purdue University, West Lafayette, IN 47907, USA

Geballe Laboratory for Advanced Materials, Stanford University, Stanford, California 94305, USA

Institute for Research in Electronics and Applied Physics, University of Maryland, College Park, Maryland 20742, USA

This document provides supplementary information to "Ultrathin and Multicolor Optical Cavities with Embedded Metasurfaces". More information includes experimental results on the nano-cavities with 240-nm polymer layer which can support high order cavity resonance determined by the Fabry-Pérot interferometric principle and the simulation data for dualband nanocavity to clarify the characteristic of observed resonance peaks in experiment.

Figure S1 shows transmission spectrum of the nanocavity with two orthogonal polarizations of incident wave. When the polymer thickness is 240 nm, the whole cavity thickness between the two metal mirrors becomes around 280 nm. In comparison with the nanocavity with 60-nm polymer layer, the 1st order cavity resonance is red-shifted to the near infrared regime and high order ($m > 1$) resonance is appeared in the visible range to satisfy the resonance conditions described in the equation (1). Even though high order harmonic resonance is introduced by increasing the thickness of the polymer spacer, Ag metasurface causes the shift of all resonances because the induced phase shift due to metasurface ($\varphi_{ms}$) has a same impact to all orders of cavity resonances. Therefore, resonant wavelength increases as the Ag gratings width increases across the values $w = 21, 35, 45, 56$, and 71 nm.

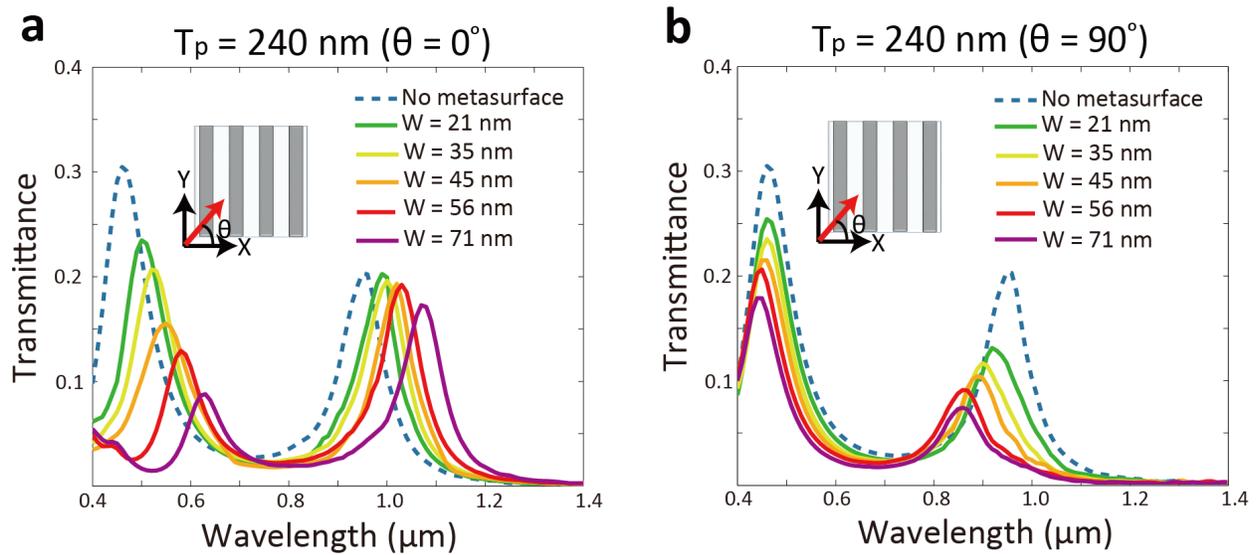

Fig S1: Experiment of transmission spectra of nanocavity with 240-nm thick polymer spacer (inset the schematic of polarization direction, θ is the angle of polarization of incident wave): (a) θ = 0° and (b) θ = 90°.

Figure S2 shows the simulation data of the nano-cavity where the metasurface is composed of alternating grating widths of 35 and 56 nm. As demonstrated in fig (3), two resonances obtained by two different grating widths agree well with the individual response of each of these two grating widths. However, in comparison with the experiment data in Fig (3), the small peak at a wavelength of 480nm is not observed in the simulation data. Considering that the simulation doesn't not include the wetting layer of Ge which strongly absorbs the light within the wavelength from 400nm to 500nm, it is obvious that the small peak at the wavelength of 480 nm in the experiment is caused by the absorption of Ge layer.

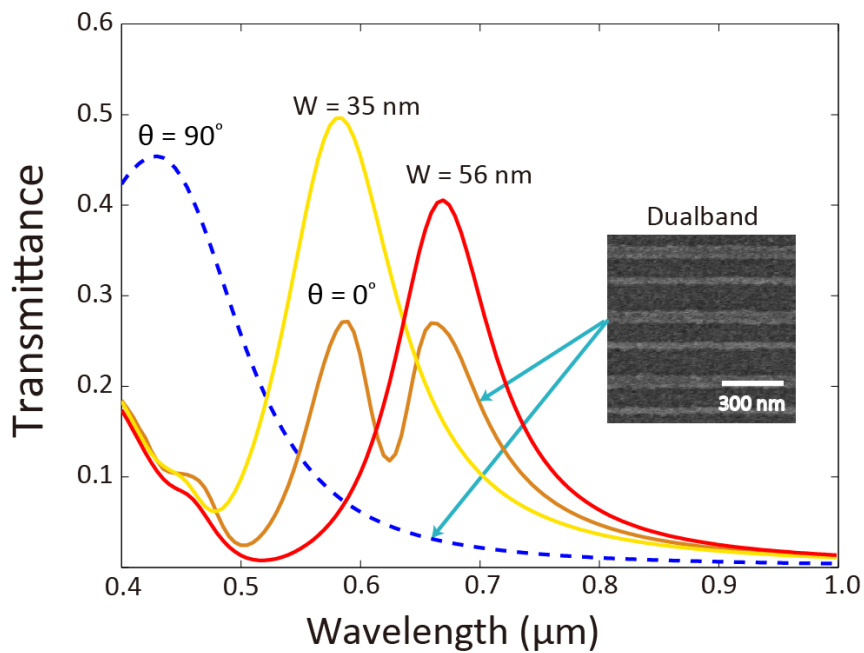

Fig S2: Simulation of transmission spectra of dualband nano-cavity with 60 nm thick polymer spacer: The dualband metasurface consists of two different Ag grating with 35 nm and 56 nm linewidth.